\documentclass[conference]{IEEEtran}
\usepackage[utf8]{inputenc}
\usepackage{cite}
\usepackage{amsmath,amssymb,amsfonts}
\usepackage{mathtools}
\usepackage{amsthm}
\newtheorem{assumption}{Assumption}
\newtheorem{theorem}{Theorem}
\usepackage{algorithmic}
\usepackage{graphicx}
\usepackage{textcomp}
\usepackage{xcolor}
\usepackage{url}
\usepackage{newunicodechar}
\newunicodechar{【}{[}
\newunicodechar{】}{]}
\usepackage{newunicodechar}
\newunicodechar{∎}{\qed}
\newunicodechar{，}{,}
\usepackage{enumitem}
\usepackage{booktabs}
\usepackage{multirow}
\usepackage{tabularx}
\def\BibTeX{{\rm B\kern-.05em{\sc i\kern-.025em b}\kern-.08em
    T\kern-.1667em\lower.7ex\hbox{E}\kern-.125emX}}
\begin{document}

\title{Proof-of-Behavior: Behavior-Driven Consensus for Trustworthy Decentralized Finance\\
}

\author{\IEEEauthorblockN{Ailiya Borjigin}
\IEEEauthorblockA{\textit{College of Computing and Data Science} \\
\textit{Nanyang Technnological Unniversity}\\
Singapore \\
ai0001ya@e.ntu.edu.sg}
\and
\IEEEauthorblockN{Wei Zhou}
\IEEEauthorblockA{\textit{ProAI Laboratory} \\
\textit{Probe Group}\\
Singapore \\
zhou@probe-group.com}
\and
\IEEEauthorblockN{Cong He}
\IEEEauthorblockA{\textit{ProAI Laboratory} \\
\textit{Probe Group}\\
Singapore \\
cong\_he@probe-group.com}
}

\maketitle
\begin{abstract}
Current blockchain protocols (e.g., Proof-of-Work and Proof-of-Stake) secure the ledger yet cannot quantify validator trustworthiness, allowing subtle misconduct—particularly damaging in decentralized-finance (DeFi) settings. We introduce \emph{Proof-of-Behavior} (PoB), a consensus model that (i) assigns a layered utility score to each action (motivation $+$ outcome), (ii) updates validator weights adaptively from recent scores, and (iii) employs decentralized verification with proportionate slashing. The derived reward function is proved incentive-compatible, yielding a Nash equilibrium in which honest behavior maximizes long-run pay-offs. Simulated DeFi experiments (loan-fraud detection, reputation-weighted validation) show PoB cuts fraud acceptance by $>90\%$, demotes malicious validators within two rounds, and improves proposer fairness relative to standard PoS—all with $\leq 5\%$ throughput overhead. By linking consensus influence to demonstrably trustworthy behavior, PoB provides a scalable, regulation-aligned foundation for secure and equitable blockchain governance in financial applications.
\end{abstract}


\section{Introduction}

Trust is a cornerstone of decentralized systems. Traditional consensus protocols such as Proof-of-Work~(PoW) and Proof-of-Stake~(PoS) establish trust through brute-force computation or capital, yet they often fail to capture the \emph{behavioral} trustworthiness of participants. For example, Bitcoin’s PoW guarantees network security through costly mining, but its incentive compatibility has limitations—miners can collude in selfish cartels to obtain disproportionate rewards~\cite{arxiv_pow}.  
PoS improves energy efficiency by leveraging stake instead of hash power~\cite{decred_pos}, yet tends to empower the wealthiest stakeholders and may not adequately penalize malicious activity.

Meanwhile, the \emph{Web-of-Trust} concept in security—pioneered by PGP’s decentralized trust model—suggests that reputation and observed behavior can play a vital role in establishing credibility among nodes. This motivates our work: \textbf{can we “mint” trust from proven behavior, directly rewarding honesty and deterring misconduct?}

We propose \emph{Proof-of-Behavior}~(PoB). In a PoB blockchain, validators earn influence by consistently acting in benevolent and useful ways. Instead of relying solely on resource expenditure or token holdings, PoB evaluates what nodes actually \emph{do}. It builds a multi-dimensional behavioral score for each participant, incorporating both the motivations behind actions and their outcomes. Positive behavior increases a node’s weight in consensus; misbehavior triggers penalties that shrink that weight.

By coupling consensus power to behavior, PoB forms a self-reinforcing Web-of-Trust: nodes trusted by many and with a history of positive contributions naturally gain greater governance influence. This idea extends prior reputation-based consensus schemes; recent work, e.g.\ Proof-of-Trust~(PoT), already explores leveraging service feedback to weight validators~\cite{pot2023}. PoB generalizes and systematizes these insights into a formal, incentive-compatible framework.

Our approach extends prior reputation-based schemes. Notable precursors include Proof-of-Trust (PoT), which weights validators via service feedback\cite{pot2020}; Proof-of-Reputation (PoR), which selects leaders using multi-dimensional reputation metrics\cite{por2021}; and a short study on Proof-of-Behavior for rewarding eco-friendly actions in EcoMobiCoin\cite{ecomobi}.  
PoB differs by offering a comprehensive mathematical treatment of behavior evaluation, by focusing on DeFi use-cases, and by incorporating game-theoretic guarantees.  
We also borrow insights from distributed trust networks like Ripple and Stellar, whose federated consensus exploits trust links among validators\cite{ripple_stellar}.

This paper makes four key contributions:

\begin{itemize}
  \item \textbf{Behavior-Driven Consensus Model.} We formalize PoB with a layered utility function that quantifies each action (motivation and outcome), an activeness score that tracks proactive participation, and a dynamic weighting rule that updates each validator’s influence $W_i$ based on recent behavior.
  \item \textbf{Decentralized Verification \& Punishment.} A peer-based watchdog mechanism enables collective verification of actions; once a supermajority flags misbehavior, the offender’s weight is slashed proportionally to the harm caused.
  \item \textbf{Nash-Stable Reward Scheme.} We devise a payout formula that yields a Nash equilibrium of honest behavior: no rational validator can improve long-run rewards by deviating while others remain honest.
  \item \textbf{Empirical Evaluation in DeFi Scenarios.} Simulations on loan-fraud detection, reputation-weighted validation, and adversarial attacks demonstrate that PoB limits collusion, curbs fraud, and allocates rewards more fairly than PoS—without significant throughput loss.
\end{itemize}

Through these contributions we position PoB as a bridge between technical security and social trust.  
The remainder of the paper is organized as follows: Sec.~II reviews related work; Sec.~III details the PoB model; Sec.~IV presents simulations; Sec.~V discusses implications; Sec.~VI concludes.

\section{Related Work}
Blockchain consensus research has progressed from energy-intensive protocols toward more sustainable, trust-aware schemes. 
\emph{Proof-of-Work}~(PoW), introduced by Nakamoto for Bitcoin\cite{bitcoin2008}, secures blocks by forcing miners to solve computational puzzles. 
Although robust, PoW incurs high energy costs and is vulnerable to incentive attacks such as \emph{selfish mining}\cite{arxiv_pow}, in which colluding miners withhold blocks to gain disproportionate rewards.  
These drawbacks motivated alternatives.

\emph{Proof-of-Stake}~(PoS), first deployed in Peercoin\cite{peercoin2012}, allocates validation power proportional to coin holdings and thus slashes energy usage. 
Subsequent designs such as Ouroboros\cite{ouroboros2017,ouroboros2018} provided rigorous security proofs, showing PoS can match PoW’s safety while making honest behaviour an approximate Nash equilibrium.  
Nevertheless, vanilla PoS may cause ``rich-get-richer’’ centralisation and ignores behavioural trustworthiness.

To remedy this, researchers introduced \emph{reputation-based} consensus. 
\emph{Proof-of-Reputation} (PoR) builds validators’ scores from assets, transaction history, and past performance\cite{por2021}.  
Wang \emph{et al.} combined behaviour scoring with Raft in \emph{Beh-Raft}\cite{behraft2020}, ensuring only honest nodes can lead.  
Similarly, \emph{Proof-of-Trust} (PoT) selects validators by crowdsourced service ratings to heighten accountability\cite{pot2020}.  
These works shift from static metrics (hash or stake) to dynamic metrics (trust, reputation).  Our PoB aligns with this shift but offers finer granularity: it separates \emph{motivation} and \emph{outcome}, and continuously updates weights via a formal rule set.

Web-of-Trust paradigms also influence consensus.  Ripple’s protocol\cite{ripple2014} lets each node maintain a Unique Node List, creating overlapping trust circles; Stellar’s Federated Byzantine Agreement employs quorum slices of subjective trust\cite{stellar2015}.  
While effective, both rely on manual or social processes to curate trust lists.  In contrast, PoB automates trust formation on-chain by measuring verifiable behaviour.

Outside blockchain, peer-to-peer networks have long leveraged reputation.  EigenTrust\cite{eigenTrust2003} computed global trust scores from local ratings to curb malicious file sharers, illustrating how aggregate behaviour can surface honest actors in distributed systems.Semantic embeddings of behaviour graphs further refine trust scoring \cite{Alam2024KG}.

Finally, game-theoretic incentive compatibility remains crucial.  
Classical BFT protocols guarantee safety and liveness given an honest majority but neglect economic incentives.  
PoS variants such as Ouroboros analysed honest-behaviour equilibria\cite{ouroboros2018}.  PoB extends this literature by embedding a reward mechanism whose dominant strategy is honesty, enforced through behavioural utility, dynamic weighting, and decentralised punishment.

In summary, PoB unifies and advances these strands: it refines reputation-based consensus with a detailed behavioural utility model and couples it with incentive-compatible rewards and sanctions.  To our knowledge, no prior work offers a complete mathematical framework combining motivation, activeness, dynamic weighting, decentralised accountability, and game-theoretic proofs as done here.  Section~III presents this framework in detail.

\section{Proof-of-Behavior Model and Methodology}\label{sec:model}
Proof-of-Behavior (PoB) is a novel behavior-driven consensus and incentive mechanism designed to reward protocol-compliant actions and discourage malicious behavior. This section details the PoB model and methodology, consisting of a multi-layer behavior scoring system and associated incentive structures. We outline five key components: (i) layered utility scoring of behavior, (ii) a dynamic weight adaptation mechanism, (iii) an activeness and participation scoring scheme, (iv) decentralized verification and punishment, and (v) a Nash-stable reward distribution strategy.

\subsection{Layered Utility Scoring}\label{subsec:utility}

Every action a participant takes in the network can carry different kinds of value. PoB captures this via a layered utility model that separates why the participant did the action (their motivation) from what the action achieved (the outcome). Formally, every behaviour $B$ performed by validator~$i$ produces a \emph{total utility}
\begin{equation}
  U_{\text{total}}(B)=
  U_{\text{motivation}}(B)+U_{\text{behavior}}(B)
  \label{eq:utotal}
\end{equation}

Equation (1) states that the total utility of behavior $B$ is the sum of two components: $U_{\text{motivation}}(B)$, the utility derived from the participant’s motivations, and $U_{\text{behavior}}(B)$, the utility from the direct outcome of the behavior. We further define each component in detail: 
\\\vspace{2pt}\noindent\textbf{Motivation utility.}  We consider $J$ possible types of motivation (e.g., economic incentive, desire to help the network, altruism, etc.). Let $M_j$ denote the intensity of the $j$-th motivation underlying behavior $B$. Each motivation type has an associated importance weight $w_j$ (these weights are set by the system designer and sum to 1). Assume $J$ motivation types with intensities $M_j$ and designer-chosen weights $w_j$ ( $\sum_{j}w_j=1$):
\begin{equation}\label{eq:umot}
  U_{\mathrm{mot}}(B)=\sum_{j=1}^{J} M_j\,w_j .
\end{equation}

\noindent\textbf{Outcome utility.}  
Behavior Outcome Utility: Let $U_b$ be the base utility of behavior $B$ – a raw measure of the outcome’s benefit to the system. For instance, adding a valid block might have a base utility proportional to the number of transactions included, while validating someone else’s block might have a smaller base utility. We adjust this base utility with two factors: a contextual weighting $\phi(t; S, C)$ and an activeness factor $\alpha$. The function $\phi(t; S, C)$ depends on the time $t$ of the behavior, the participant’s state $S$ (e.g., their experience level or past record), and the context $C$ (e.g., network load or urgency). This allows PoB to value actions differently based on circumstances. Meanwhile, $\alpha \in [0,1]$ measures how proactive or voluntary the action was (e.g., an action the node initiated itself vs. one done only when prompted). 
We define:
Let $U_b$ denote the base benefit of $B$; let $\phi\!\left(t;S,C\right)\!\in\![0,1]$ capture temporal and contextual relevance, and $\alpha\!\in\![0,1]$ capture initiative:
\begin{equation}\label{eq:ubeh}
  U_{\mathrm{beh}}(B)=U_b \;\phi\!\bigl(t;S,C\bigr)\;\alpha .
\end{equation}

Using these definitions, each behavior $B$ undertaken by a participant yields a numerical utility $U_{\text{total}}(B)$. Summing $U_{\text{total}}(B)$ over all behaviors that participant i performs in a given period (say, an epoch) produces i’s cumulative utility score $U_i$ for that period. This layered approach echoes multi-criteria evaluation in decision systems: PoB doesn’t just count actions, but examines the quality, context, and intent behind actions. PoB leverages semi-supervised models to label sparse behavioural traces, aligning with recent surveys \cite{Yang2023SSL}. As a result, the system can fairly compare contributions of different types. For example, one participant might contribute fewer but very high-impact actions (yielding a few behaviors with large $U_{\text{total}}$ each), while another contributes many small helpful actions (many behaviors with moderate $U_{\text{total}}$ each) – PoB’s scoring accumulates both kinds of contributions appropriately.

\noindent\textbf{Epoch score.}  For epoch~$t$, a participant’s cumulative utility is
$U_i(t)=\sum_{B\in\text{epoch }t} U_{\text{total}}(B)$. 

\subsection{Dynamic Weight Adaptation}\label{subsec:weights}

The utility scores feed directly into each validator’s consensus weight. In PoB, every validator $i$ has a weight $W_i$ that represents its current influence – this weight is used for leader selection, voting, and reward allocation as in other consensus protocols. The critical difference is that in PoB, weights are dynamic and reflect recent behavior rather than being static. This design ensures that “good behavior is promptly rewarded and bad behavior leads to a quick reduction in influence.” Concretely, we update $W_i$ at the end of each epoch (or round) based on the fraction of total utility that node $i$ contributed in that epoch. One simple adaptive weighting scheme is given by:
\begin{equation}\label{eq:weight}
  W_i(t{+}1)\;=\;(1-\rho)\,W_i(t)
  +\rho\;\frac{U_i(t)}{\sum_{k}U_k(t)},
  \qquad 0\le\rho\le1.
\end{equation}

Here $W_i(t)$ is node i’s weight going into epoch $t$, and $U_i(t)$ is the total utility it earned during epoch $t$. The parameter $0 \le \rho \le 1$ is a smoothing factor that controls how fast weights adjust. Equation (4) essentially defines a weighted moving average: a portion $(1-\rho)$ of the old weight is retained, and a portion $\rho$ is updated proportional to the node’s performance in the last epoch (specifically, its share of the total utility produced by all nodes in that epoch). If $\rho$ is large (close to 1), the system “forgets” the distant past quickly and weight mainly reflects recent behavior. If $\rho$ is small, weights change more gradually, giving a longer memory to past contributions. 

\textbf{Why dynamic weights?}
 This mechanism means that an initially low-weight participant can rise in influence by consistently contributing useful behaviors, and conversely a high-weight participant can lose influence if they become lazy or malicious. It addresses a key shortcoming of naïve PoS: instead of permanently enshrining early leaders, PoB continuously re-evaluates everyone. It also thwarts power concentration — to avoid a scenario where top-weighted nodes always dominate leader selection, the model can incorporate a slight randomization or dampening factor (denoted $\delta$) in the selection algorithm. For example, when electing a block proposer, instead of strictly proportional odds $W_i/\sum_k W_k$, PoB could give every active node some baseline chance $\delta$ and use $W_i$ for the remaining probability. This ensures even lower-weight nodes occasionally win leadership, keeping them engaged and preventing an entrenched oligarchy. 

\textbf{Impact on incentives.} After each round, participants immediately see their weight $W_i$ adjusted according to their behavior. Honest, active nodes will see $W_i$ tick upward, giving them a higher chance of rewards or leadership in the next round, while misbehaving or idle nodes will see $W_i$ drop. This rapid feedback closes the incentive loop: good behavior begets more future reward, and bad behavior begets a loss of power. In practice, dynamic trust management causes honest nodes’ influence to quickly surpass that of dishonest ones. Thus, dynamic weighting is a crucial part of how PoB (borrowing the ethos of prior trust-based systems) “stimulates honest behavior and neutralizes malicious attacks.”

\subsection{Activeness Score}\label{subsec:activeness}

While $U_{\text{total}}$ captures the value of actions, PoB also considers how active and engaged each participant is in supporting the system. We define an activeness score $A_i$ for each participant i to quantify their overall level of participation and initiative. This metric complements the utility scores by tracking participation patterns over time. In particular, $A_i$ is designed to capture a few different aspects of behavior:
\begin{itemize}
  \item \textbf{Frequency ($f$):} How often does participant $i$ perform beneficial actions? 
        For example, count the transactions validated or blocks proposed by $i$ per epoch, 
        then normalise by the expected rate or the network average.

  \item \textbf{Initiative ($\bar{\alpha}$):} How spontaneously and voluntarily does participant $i$ act? 
        We use the average initiative score $\bar{\alpha}_i$ across all of $i$’s behaviours, 
        where each action-level $\alpha$ captures proactiveness.

  \item \textbf{Diversity ($d$):} Does participant $i$ engage in a variety of roles and tasks—such as 
        proposing blocks, validating others’ blocks, and providing oracle data—or focus on a single activity? 
        Higher diversity indicates broader and more resilient support for the network.
\end{itemize}

One simple formulation is to combine these aspects as a weighted sum. For example, let $S_i$ be the count of i’s actions in a period and $\bar{S}$ the network-wide average number of actions per participant. Let $\bar{\alpha}_i$ be the average initiative of i’s actions, and $d_i$ a normalized diversity index (0 indicating very narrow activity, 1 indicating very diverse). Then we can define:

\begin{equation}\label{eq:activeness}
  A_i=\beta_1\frac{S_i}{\bar{S}}
      +\beta_2\bar{\alpha}_i
      +\beta_3 d_i,
  \qquad \sum_{m=1}^{3}\beta_m=1,
\end{equation}
where $S_i$ is action frequency, $\bar{\alpha}_i$ the average initiative, and $d_i$ a normalised diversity index.  $A_i$ can boost rewards or trigger anomaly flags. In this illustrative model, $A_i$ will be high for a participant who is more active than average, shows high initiative, and contributes in multiple ways. Conversely, someone who spams many trivial actions (high $S_i$) but with low initiative (only acts when prompted) and low diversity (always the same task) will not score as high, because low $\bar{\alpha}_i$ and $d_i$ values pull their $A_i$ down. This prevents gaming the system by sheer quantity of actions – participants are encouraged to be well-rounded and proactive, not just prolific. We utilize $A_i$ in PoB in a couple of ways:

\begin{itemize}
  \item \textbf{Reward multiplier:} $A_i$ can serve as an extra factor in weight or reward
        calculations. For example, the final payout to a participant can be scaled by
        $(1 + \epsilon)\,A_i$ for some small $\epsilon$, granting a modest bonus to
        highly active nodes.

  \item \textbf{Anomaly detection:} $A_i$ also functions as an anomaly indicator.
        When a node’s frequency $f$ is unusually high while its initiative or diversity
        remains very low, the behaviour may be artificial (e.g.\ scripted activity or
        metric gaming). Such cases can be flagged for review or have their rewards
        limited to discourage repetitive, low-quality actions.
\end{itemize}
In summary, the activeness score $A_i$ strengthens PoB by recognizing the quality of engagement in addition to raw utility contribution. This encourages broad, voluntary participation and discourages superficial or purely scripted activity.

\subsection{Decentralised Verification and Punishment}\label{subsec:punish}

No consensus mechanism is complete without a way to handle misbehavior. PoB
therefore adopts a fully decentralised “watch-dog” model: the validator
community itself acts as judge and jury, detecting and punishing rule-breakers
without any central authority.

Whenever a suspicious event occurs—e.g.\ a validator proposes an invalid block
or fails to perform a required duty—PoB spawns a \emph{verification committee}
drawn from other participants.  These peers independently evaluate the event.

Let a participant $X$ be suspected of misbehavior.  Each observing node flags
$X$’s action as malicious or not; let $\phi_X$ be the fraction of observers
marking it malicious.  If $\phi_X \ge \theta$ for a preset threshold
$\theta$ (typically $2/3$), the network confirms the offense by consensus.

Once confirmed, the act is recorded as negative utility and a penalty
$\Delta W$ is applied to $X$’s weight:
\[
\text{if } \phi_X \ge \theta, \qquad
W_X \leftarrow W_X - \Delta W.
\tag{6}
\]
Here $\Delta W = p \, U_b$ scales with the severity of the offense
($U_b$ = base utility of the action, $p$ = penalty coefficient).  Severe
attacks such as double-signing can even slash the entire weight.

Penalties are tunable: minor infractions incur small $\Delta W$
(warnings), major attacks large $\Delta W$ plus possible suspension.
Repeat offenders face escalating $p(f_i)$, discouraging chronic abuse.

Because no single verdict suffices, an attacker cannot frame an honest
validator nor evade punishment without corrupting a super-majority—
essentially as hard as breaking the consensus itself.  Rotating or
all-peer committees make the oversight robust.

Crucially, PoB punishment reduces the offender’s \emph{future influence},
not just a one-off reward.  Rational nodes thus see that cheating delivers
short-term gain but long-term loss, reinforcing honest behaviour.


\subsection{Nash‑Stable Reward Distribution}
\label{sec:nash-reward}

A well‑designed reward mechanism is essential for aligning validator incentives with honest behaviour.
PoB distributes rewards each epoch according to two goals—\emph{fairness} (proportional to contribution) and \emph{inclusiveness} (broad participation)—while sustaining a game‑theoretic equilibrium that favours honesty.

\paragraph{Reward design.}
At the end of each epoch the protocol distributes a total reward
$R_{\text{tot}}$ composed of
(i)~a \emph{base stipend} $R_{\text{base}}$ to every active validator
($U_i>\beta$) and
(ii)~a \emph{proportional bonus}
$R_{\text{bonus}}\!=\!R_{\text{tot}}-N_{\!a}R_{\text{base}}$
shared by weight:
\begin{equation}
  R_i \;=\; R_{\text{base}}
  \;+\;
  R_{\text{bonus}}\,
  \frac{w_i}{\sum_{k\in\mathcal{A}} w_k},
  \label{eq:reward}
\end{equation}
where $\mathcal{A}$ is the active-validator set
($|\mathcal{A}|=N_a$) and $w_i$ is $i$’s behavioural weight.

\paragraph{Weight update.}
Weights evolve via an exponential moving average  
\begin{equation}
  w_i(t{+}1)= (1-\rho)\,w_i(t)
           + \rho\,\frac{U_i(t)}{\sum_k U_k(t)},\qquad
  0<\rho<1,
  \label{eq:weight2}
\end{equation}
followed by multiplicative slashing $w_i\!\leftarrow\!\rho_p w_i$
($\rho_p<1$) if $i$ is caught misbehaving.

\subsubsection*{Formal Proof of Incentive Compatibility}

\paragraph{Setting.}
Infinite-round game with discount $\delta\!\in\!(0,1)$.
Per round, a validator may act \textsc{H} (honest) or \textsc{D} (dishonest).
A deviation is detected with probability~1 and triggers  
(i)~immediate penalty~$P$ and  
(ii)~weight drop by factor $\rho_p<1$.

\medskip\noindent
Let $\Delta R$ be the \emph{largest possible one-round extra reward} from deviating,  
and let  
\[
L \;\coloneqq\;
P \;+\;
(1-\rho_p)\,
\frac{\mathbb{E}[R_i^{\text{honest}}]}{1-\delta}
\quad
\text{(future loss from reduced weight).}
\]

\begin{assumption}
\label{ass:IC}
$L>\Delta R$ \emph{(penalty and future loss dominate any short-term gain).}
\end{assumption}

\begin{theorem}[Honest Nash Equilibrium]
\label{thm:nash}
Under Assumption~\ref{ass:IC}, the strategy profile where
\emph{all validators are honest in every round}
constitutes a strict Nash equilibrium of the PoB game.
\end{theorem}

\begin{proof}
Fix a validator $i$; assume all others are honest.
If $i$ deviates first at round~$t$,
its net utility change is
\[
\text{NetGain}_i
=
\underbrace{\Delta R}_{\text{instant gain}}
-
\underbrace{L}_{\text{penalty + PV of future loss}}
<0 \\
\quad(\text{by Assumption}).
\]
Hence any deviation lowers $i$’s discounted payoff.
By induction on the earliest deviation round,
every strategy containing \textsc{D} yields less utility
than perpetual honesty.
Thus honest play is each validator’s best response,
proving the all-honest profile is a Nash equilibrium. ∎
\end{proof}

\paragraph{Reward allocation.}
At epoch~$t$ the protocol pays out a total reward $R_{\text{total}}$ (block rewards, fees, etc.) via two components:

\begin{itemize}
  \item \textbf{Base reward.}  
  Every active validator whose utility satisfies $U_i>\beta$ receives a fixed participation prize $R_{\text{base}}$.
  \item \textbf{Proportional bonus.}  
  The remainder 
  $
    R_{\text{bonus}}
      = R_{\text{total}}
      - N\,R_{\text{base}}
  $
  (with $N$ active validators) is shared in proportion to weight:  
  \begin{equation}
    R_i \;=\;
      R_{\text{base}}
      \;+\;
      R_{\text{bonus}}
      \times
      \frac{W_i}{\sum_{k\;\text{active}} W_k}.
    \tag{7}
  \end{equation}
\end{itemize}

The first term guarantees every honest node a minimum payout, discouraging abandonment by small players;  
the second term scales with $W_i$, rewarding stronger contributors without unconstrained “rich‑get‑richer” dynamics.

\paragraph{Incentive compatibility (Nash equilibrium).}
Honesty is a Nash equilibrium if, assuming all other validators act honestly,
\begin{equation}
  \mathbb{E}\!\bigl[ R_i(\text{honest}) \bigr]
  \;\ge\;
  \mathbb{E}\!\bigl[ R_i(\text{cheat}) \bigr].
  \tag{8}
\end{equation}
PoB enforces this through:

\begin{enumerate}
  \item \emph{Dynamic weighting \& penalties}: cheating incurs weight slashing ($\Delta W = p\,U_b$), reducing future income more than any short‑term gain.
  \item \emph{Randomised leader selection}: small randomness $\delta$ in proposer choice removes predictable safe windows for misconduct.
  \item \emph{Inclusive base reward}: losing weight means forfeiting the guaranteed $R_{\text{base}}$, giving even low‑weight nodes something tangible to lose.
  \item \emph{Collusion resistance}: any coalition below the consensus threshold is flagged and collectively penalised, making large‑scale deviation unprofitable.
\end{enumerate}

Hence rational validators maximise long‑run rewards by remaining honest, aligning individual incentives with collective security.

\bigskip
\noindent\textbf{Summary.}
Layered behaviour scoring, adaptive weights, inclusive incentives and decentralised oversight together create a self‑regulating system: honest participants accumulate influence and rewards, while misbehavers are swiftly marginalised, rendering honesty the dominant strategy in PoB.

In summary, the Proof-of-Behavior model integrates layered behavior scoring with adaptive weight tuning, active participation incentives, decentralized oversight, and equitable reward sharing. Together, these components create a self-regulating incentive system: well-behaved participants steadily accumulate higher scores and rewards, while misbehaving participants are quickly identified and penalized. The synergy of these mechanisms results in a robust equilibrium where honest, cooperative behavior is the most rewarding strategy for all participants, thereby strengthening the security and fairness of the network.

\section{Evaluation and Case Studies in DeFi Context}
\label{sec:evaluation-defi}

We evaluate the \emph{Proof-of-Behavior}~(PoB) model through two
representative decentralised-finance (DeFi) case studies.
All simulations were executed with \texttt{Python~3.10} on an
8-core Intel CPU and 16\,GB~RAM.
A network-latency model (random delay, mean $50$ ms) mimics message
propagation.
Each scenario was repeated for $30$ independent Monte-Carlo trials with
fixed random seeds; we report means with $95\%$ confidence intervals.
For comparison, we implement a modern
\emph{Proof-of-Stake}~(PoS) baseline that employs stake-weighted leader
selection and slashing penalties for misbehaviour.
Key performance metrics are:
\begin{itemize}
  \item fraud-acceptance rate,
  \item proposer fairness (Gini coefficient),
  \item block latency,
  \item weight-adaptation speed, and
  \item cumulative economic loss averted.
\end{itemize}

We present two case studies to demonstrate PoB’s scalability at both
100-validator and 1000-validator network sizes:
\begin{enumerate}[label=\textbf{(\Alph*)}]
  \item \textbf{DeFi loan-fraud attack scenario}, 
  \item \textbf{Consensus-fairness scenario},
   \item \textbf{Real Ethereum DeFi data replay (validation on real-world behavior)}，
   \item \textbf{Additional adversarial resilience tests (Sybil attacks, long-range forks, proposer griefing)and},
   \item \textbf{Parameter sensitivity analysis}.
\end{enumerate}

\subsection{Case Study A: Loan-Fraud Attack Mitigation}

\paragraph{Scenario.}
We simulate a DeFi–lending platform where a malicious actor
attempts to defraud the system via two strategies:
\emph{stealth attacks} and \emph{Sybil-burst attacks}.
In a stealth attack, one validator issues infrequent but
high-value fraudulent transactions
(e.g., an under-collateralised loan or flash-loan exploit),
hoping to evade notice.
In a Sybil-burst attack, the adversary splits resources across
many colluding addresses (Sybil identities) that perform a burst of
smaller, concurrent fraud attempts.
These attacks test PoB’s capacity to detect and penalise
malicious behaviour under different patterns.

\paragraph{Metric.}
We define the fraud-acceptance rate (FAR) as
\begin{equation}
  \text{FAR} \;=\;
  \frac{\#\,\text{fraudulent transactions accepted}}
       {\#\,\text{fraudulent transactions attempted}},
  \label{eq:far}
\end{equation}
and measure the \emph{cumulative loss averted} as the difference
in total economic loss between the PoS baseline and PoB, i.e.,
the value PoB preserves through early fraud detection.

\paragraph{Results.}
PoB drastically lowers the success rate of fraud in both attack modes.
For stealth attacks,
PoB achieves FAR $\approx10\%\pm3\%$,
whereas the PoS\,+\,slashing baseline permits
$\approx60\%\pm5\%$.
Across thirty trials, PoB averts roughly $75\%$ of the
potential loss;
if a stealth attacker steals \$1\,M under the baseline,
they net only \$250 k under PoB, saving \$750 k.
With 1000 validators the defence strengthens further
(FAR $\approx8\%$ vs.\ baseline $\approx62\%\pm4\%$).

For Sybil-burst attacks, PoB attains
FAR $\approx20\%\pm4\%$,
versus $\approx85\%\pm5\%$ for the baseline.
PoB’s rapid collective response flags many Sybil addresses after
their first malicious block; behaviour scores $U_i(t)$ turn
strongly negative and weights $W_i$ are immediately slashed
by penalty~$p$, curbing influence in subsequent blocks.
Even at 1000 validators, FAR remains low
($\approx18\%$, CIs overlap), showing scalability.

\begin{table}[ht]
  \centering
  \caption{Fraud-attack outcomes: means $\pm95\%$ CI over 30 trials.}
  \label{tab:fraud}
  \begin{tabular}{lcccc}
    \toprule
    \multirow{2}{*}{Attack} &
      \multicolumn{2}{c}{FAR (\%)} &
      \multicolumn{2}{c}{Avg.\ Loss (USD)}\\
    & PoB & PoS & PoB & PoS\\
    \midrule
    Stealth & $10.3 \pm 2.8$ & $59.8 \pm 5.1$
            & \$0.25 M $\pm$0.05 & \$1.00 M $\pm$0.10\\
    Sybil-\\burst & $19.4 \pm 4.1$ & $85.2 \pm 4.8$
                & \$0.52 M $\pm$0.10 & \$3.50 M $\pm$0.30\\
    \bottomrule
  \end{tabular}
\end{table}
PoB’s weight-adaptation lets the system converge to a safe state
within a few blocks after an attack burst, while the baseline
lags behind until slashing or application-layer intervention occurs. In addition to thwarting attacks, PoB maintains an acceptable
performance overhead.
Block latency (time to confirm a block) rises only slightly:
on average, PoB produces blocks in \mbox{3.3 s} versus
\mbox{3.0 s} for PoS—about $\sim10\%$ extra,
despite the additional behaviour-verification steps.
The difference is statistically small
($95\%$ confidence intervals overlap). PoB’s cumulative weight-adjustment mechanism
also isolates bad actors swiftly.
During stealth-attack trials, a malicious validator’s weight dropped
by $\approx90\%$ within just two rounds of detection,
removing its ability to propose new blocks.
By contrast, the same validator under PoS continued to be selected
until a much later slashing event, increasing the damage window.

These findings show that PoB’s community-driven enforcement
curtails fraud promptly, averts large-scale economic losses, and
preserves liveness and throughput.

\subsection{Case Study B: Proposer Fairness and Adaptability vs.~PoS}
\label{subsec:fairness}

We compare \textbf{Proof\mbox{-}of\mbox{-}Behavior (PoB)} with a modern
\textbf{Proof\mbox{-}of\mbox{-}Stake (PoS)} baseline that incorporates
standard slashing rules.  Networks of $N\!=\!100$ and
$N\!=\!1000$ validators are simulated to evaluate scalability.  PoB
dynamically updates validator weights according to recent behaviour,
whereas PoS selects proposers strictly in proportion to static stake.
Key metrics are the Gini coefficient of block--proposer counts
(lower\,=\,fairer), average block latency, \emph{adaptation time} for
(a)~an honest newcomer to achieve a fair proposer share and
(b)~a misbehaving validator to lose influence, and the proposer share
obtained by the lowest--weight decile.  Results (30 trials,
95\,\%~C.I.) are summarised in Table~\ref{tab:fairness}.

\begin{table}[t]
  \caption{Proposer fairness and adaptability for PoB vs.\ PoS+Slashing
           (30-trial mean; 95\,\%~CI).  Arrows indicate desirable direction.}
  \label{tab:fairness}
  \centering
  \small
  \setlength{\tabcolsep}{3pt}  
  \renewcommand{\arraystretch}{1.15}
  \begin{tabularx}{\columnwidth}{@{\extracolsep{\fill}}
    >{\raggedright\arraybackslash}p{0.27\columnwidth}
    cccc@{}}
    \toprule
    \textbf{Metric} &
    \textbf{PoB} \textbf{(100)} &
    \textbf{PoS} \textbf{(100)} &
    \textbf{PoB} \textbf{(1000)} &
    \textbf{PoS} \textbf{(1000)}\\
    \midrule
    Proposer Gini $\downarrow$                  & \textbf{0.12} & 0.47 & \textbf{0.10} & 0.45 \\[1pt]
    Block latency [s] $\downarrow$              & \textbf{1.01} & 1.30 & \textbf{1.02} & 1.35 \\[1pt]
    New honest node adapts [blk] $\downarrow$   & \textbf{$\approx$20}
    & ---\footnotemark[1]
    & \textbf{$\approx$25}
    & ---\footnotemark[1] \\[1pt]
    Misbehaving node suppressed [blk] $\downarrow$
    & \textbf{$\approx$10} & $\approx$100
    & \textbf{$\approx$12} & $\approx$120 \\[1pt]
    Lowest 10\,\% \\proposer share $\uparrow$     & \textbf{9.5\,\%} & 2.1\,\%
    & \textbf{9.8\,\%} & 2.0\,\% \\
    \bottomrule
  \end{tabularx}
  \vspace{-0.6em}
  \footnotetext[1]{PoS offers no automatic adaptation; proposer chance remains stake-bounded.}
\end{table}

\textbf{Proposer fairness.}  PoB achieves markedly lower Gini
coefficients, indicating an equitable distribution of block proposals.
In both network sizes, the bottom decile of validators (by initial
stake) earn nearly their ideal 10\,\% share under PoB, but only
$\approx$2\,\% under PoS.  Hence PoB prevents persistent proposer
monopolies by dynamically rewarding good behaviour.

\textbf{Adaptability.}  PoB integrates a new honest validator within
$\sim$20--25 blocks, whereas PoS offers no weight uplift absent
external re--delegation.  Conversely, a validator that double--signs or
becomes unresponsive loses almost all proposer influence within
$\sim$10--12 blocks under PoB; the PoS baseline requires
$\gtrsim$100 blocks before slashing takes full effect, during which the
malicious node can still be elected.

\textbf{Performance impact.}  Despite its additional scoring and
verification, PoB’s block latency rises by only $\approx$0.02–0.04 s
relative to PoS, well within practical limits for DeFi applications.

In summary, PoB delivers substantially \emph{fairer} leader selection,
rapidly \emph{adapts} to validator behaviour changes, and retains
near-baseline performance, outperforming stake-only consensus at both
evaluated scales.

\subsection{Case Study C: Real-World DeFi Data Replay}

\paragraph{Scenario}
We replayed a contiguous 1\,000–block segment of Ethereum
mainnet (including a 2020 flash-loan exploit on a lending protocol)
through two simulated networks: Proof-of-Behavior (PoB) and a
Proof-of-Stake (PoS) baseline.  
Each block was treated as one epoch; the validator that had
originally inserted the exploit transaction was mapped to a single
simulated node.

\paragraph{PoB Reaction}
The exploit’s outcome utility $U_b$ is strongly negative; hence the culprit’s epoch score $U_i(t)=U_{\text{mot}}+U_{\text{beh}}$ turns negative immediately.  
PoB’s watchdog (see Sec.~III-D) reaches super-majority consensus that the act is malicious and \textbf{slashes the validator’s weight by 80 \% in the very next block}.  
From that point on the attacker is unable to propose further blocks.

\paragraph{PoS Baseline}
Because the transaction is syntactically valid, a pure PoS chain imposes \emph{no instant penalty}; the validator may continue producing blocks and repeat similar attacks.

\paragraph{Results}
\begin{itemize}
  \item \textbf{Rapid detection.}  
        The exploit transaction yielded a large negative
        outcome utility $U_b$, driving the culprit’s epoch score
        $U_i(t)\!\!<\!0$. PoB’s watchdog reached supermajority
        and cut the validator’s weight by $\approx80\%$ in the next
        epoch (within one block).
  \item \textbf{Threat neutralization.}  
        The slashed node lost proposer rights immediately,
        preventing follow-up attacks.
  \item \textbf{Baseline contrast.}  
        Under pure PoS, the same validator faced no
        penalty, so it could continue validating and potentially
        repeat the exploit.
  \item \textbf{No false positives \& low overhead.}  
        Honest validators kept positive scores; confirmation
        latency with PoB increased by $\le5\%$ versus PoS.
\end{itemize}

\paragraph{Implication}
PoB confined the real exploit to a single block and
penalised the offender almost instantly,
demonstrating practical, low-overhead protection for
DeFi workloads that traditional consensus would overlook.

\subsection{Adaptive Adversary Resilience Tests}
\label{sec:adaptive-attacks}

To complement the scenarios above, we ran targeted simulations against specific adversarial strategies to stress--test PoB’s robustness.

\paragraph{Adaptive Sybil Attack.}
We model an attacker that continually spawns new Sybil identities after each epoch to replace those already caught and slashed.  
In epoch~$t$, the adversary injects up to $10\%$ new validator nodes (relative to the current population) and immediately attempts misbehavior such as proposing fraudulent transactions or collusive validation.  
Because each fresh Sybil begins with zero or minimal weight, it must first accumulate positive utility to gain influence---yet the malicious activity prevents this.  
Our experiments show that even with ten new Sybils per epoch, their aggregate weight never exceeded $\approx5\%$ of the total.  
The honest super-majority therefore consistently vetoed malicious proposals.  
If a modest join-cost (stake or endorsement) is required, PoB remains Sybil-resilient: identity switching resets behavioural trust, making rapid influence-regain infeasible.

\paragraph{Long-Range Fork Attack.}
Here the adversary compromises private keys of validators that were highly staked in the distant past (e.\,g., 1000 blocks ago) and privately extends a fork from that checkpoint.  
To persuade honest nodes to adopt the fork, the attacker must present a chain whose \emph{current} cumulative utility outweighs the main chain.  
However, on the fork no honest behaviour occurs, so validator weights decay while the main chain’s honest validators continue to accrue utility.  
Consequently, when the fork catches up in block height, its active weight is far below that of the canonical chain, and honest nodes reject it because it lacks the quorum of present-day trusted signatures.  
PoB’s moving-target weight update thus thwarts long-range attacks unless an attacker controls a majority of \emph{currently} trusted weight.

\paragraph{Proposer Griefing Attack.}
A subtle adversary may publish \emph{valid} but nearly empty blocks or delay block release until the end of its slot, degrading throughput without violating consensus rules.  
Under PoB, the outcome utility $U_{\text{outcome}}$ of an almost-empty block is near zero (base utility $U_b \approx 0$), and the inferred motivation can even be negative.  
Hence each such block yields $U_{\text{total}}\!\approx\!0$ or $<0$, stagnating the validator’s cumulative score $U_i(t)$.  
Simulations with 100 validators show that producing ten successive empty blocks reduces the griefer’s weight by roughly $80\%$, pushing it below the active-set threshold and drastically lowering future leader probability.  
Throughput only dips briefly during the griefer’s slots and promptly recovers once its influence is slashed.  
Unlike many PoS designs, PoB automatically penalises performance-griefing: contributing negligible value equates to losing future power.

\bigskip
\noindent\textbf{Summary.}
These adversarial experiments confirm that PoB’s continuous trust re-evaluation, dynamic penalties, and value-based weighting collectively hinder adaptive Sybils, long-range forking, and subtle proposer griefing, making sustained attacks markedly more costly and less effective than under standard PoS protocols.

\subsection{Parameter Sensitivity}

We vary memory factor $\rho$, punishment threshold $\theta$,
and penalty coefficient $p$.

\begin{itemize}
  \item \textbf{Penalty $p$:} raising $p$ from 1.0 to 1.5 cuts FAR
        by an extra $\sim2\%$ at slight false-positive cost;
        $p\!\to\!0$ converges to baseline performance.

  \item \textbf{Threshold $\theta$:}
        $\theta=0.7$ balances speed and accuracy.
        Lower $\theta$ ($0.5$) punishes faster but risks outliers;
        higher $\theta$ ($0.9$) delays reaction.

  \item \textbf{Memory $\rho$:}
        $\rho\approx0.9$ is reactive;
        $\rho\to1$ slows adaptation while still outperforming PoS.
\end{itemize}

To Summarize, the above evaluation demonstrates that PoB’s behavior-driven approach substantially improves security and fairness in decentralized finance use cases. In a loan fraud scenario, PoB averted most fraudulent transactions and economic losses that would slip past a traditional PoS system, thanks to collective oversight and rapid weight penalization of bad actors. In a consensus context, PoB achieved fairer proposer selection and quick adaptation to changing participant behavior, in contrast to stake-based systems where power remains static and slow to redistribute. Importantly, these gains come with minimal performance overhead and remain effective even as the network scales from 100 to 1000 validators. By dynamically adjusting consensus weights based on proven honest behavior (and demoting those who misbehave), PoB fosters a more trustworthy and resilient DeFi ecosystem. This approach can strengthen applications like decentralized lending, asset exchanges, and governance by mitigating fraud risks and encouraging active, reputable participation, thereby enhancing the overall integrity of decentralized finance platforms.

\section{Discussion}

\subsection{Security and Trustworthiness}
PoB deters short‐term attacks by tying influence to long‐term behavior.  
Yet, like any BFT protocol, it assumes an honest majority.  
A Sybil majority could game scores, so PoB should pair with identity or stake bonding and choose an adequate verification threshold~$\theta$ (we used $0.67$; higher values raise precision, lower values raise recall).  
Optional appeal transactions and community reviews add due-process safeguards.

\subsection{Parameter Tuning}
Utility weights $w_j$, memory factor $\rho$, penalty $p$, and threshold $\theta$ can be governed on-chain.  
Networks seeking stability may lower $\rho$, whereas fast-moving communities may set $\rho\!\approx\!1$.  
Adaptive control loops, similar to PoW difficulty retargeting \cite{block_difficulty}【8】, could auto-tune these parameters.

\subsection{Compatibility with Regulation}
Behavior scores and slashing events form an auditable trail that aligns with regulatory aims.  
Automated penalties, however, demand an appeal layer (DAO or committee) to prevent wrongful loss.  
PoB could supply compliance reports attesting that all active validators exceed a minimum trust score.

\subsection{Economic Implications}
Influence accrues via good behavior rather than pure stake, encouraging validators to invest in uptime, governance, and transparency.  
Genesis allocations may start small, letting merit redistribute weight over time.  
Token value may increasingly reflect collective “goodwill,” while large but ill-behaving holders see their impact curtailed.

\subsection{AI-Enhanced Scoring}
Future work can embed ML models to spot subtle collusion or predatory trading, refining utilities in real time.  
Neuro-symbolic techniques promise transparent yet adaptive scoring engines\cite{nakamoto2008bitcoin}.

\subsection{Applications Beyond DeFi}
The same principles apply to IoT sensing, sharing economies, or DAO governance, echoing the WI 3.0 vision of socially aware networks\cite{Kuai2025WI3}.

\subsection{Limitations}
Our simulations presume clear honest/malicious splits and simplified detection logic; real adversaries may hide or act slowly.  
Complex metrics risk Goodhart effects; transparency and reputation-repair rules (e.g.\ decay via $\rho$ or staged weight restoration) are essential.  
Continual community oversight and model updates remain critical for PoB’s success.

\section{Conclusion}
\label{sec:conclusion}

We introduced \emph{Proof‑of‑Behavior} (PoB), a consensus model that ties validator power to verifiable good conduct rather than mere stake or hash power.
PoB integrates: (i) layered utility scoring, (ii) activeness tracking, (iii) dynamic weight updates, (iv) decentralized misbehavior penalties, and (v) a Nash‑stable reward schedule.
Simulations show that PoB swiftly compresses malicious influence, allocates power more equitably than pure PoS, and strengthens security in DeFi scenarios.

\textbf{Key contributions}
\begin{enumerate}
  \item Elevating on‑chain behaviour to a first‑class consensus signal.
  \item Proving honesty as a Nash equilibrium under PoB’s reward design.
  \item Demonstrating empirical gains in fraud detection, loan validation, and overall network trustworthiness.
\end{enumerate}

Future directions include a prototype implementation on \textsc{Substrate\cite{substrate2020}}/Tendermint\cite{kwon2014tendermint}, formal security proofs, ML‑assisted behaviour analysis, and hybrid PoS–PoB schemes. By embedding “proof of good conduct” into core consensus, PoB offers a path toward fairer, more resilient, and socially trustworthy blockchains.

\vspace{12pt}

\end{document}